\documentclass[modern]{aastex63}
\usepackage[utf8]{inputenc}
\usepackage{soul}
\usepackage{xcolor}
\usepackage{placeins}
\usepackage{ltablex}
\usepackage{longtable}
\usepackage{graphicx}

\usepackage{graphicx,booktabs}

\usepackage{float}

\bigskip
\accepted{19 January 2020}
\submitjournal{The Astronomical Journal}
\shorttitle{29P/SW1 gas and dust during outbursts}
\shortauthors{Wierzchos and Womack}
\begin{document}
\title{\bf{CO gas and dust outbursts from Centaur 29P/Schwassmann-Wachmann}}
\correspondingauthor{K. Wierzchos}
\email{kacperwierzchos@gmail.com}

\author[0000-0002-0786-7307]{K. Wierzchos}
\affiliation{Catalina Sky Survey \\
Lunar and Planetary Lab \\
University of Arizona \\
Tucson, AZ, 85721 USA}

\author[0000-0003-4659-8653]{M. Womack}
\affiliation{Florida Space Institute \\
Department of Physics\\
University of Central Florida \\
Orlando, FL USA}

\begin{abstract}

29P/Schwassmann Wachmann is an unusual solar system object. Originally classified as a short-period comet, it is now known as a Centaur that recently transferred to its current orbit, and may become a Jupiter Family comet. It has exhibited a dust coma for over 90 years, and regularly undergoes significant dust outbursts. Carbon monoxide is routinely detected in high amounts and is typically assumed to play a large role in generating the quiescent dust coma and outbursts. To test this hypothesis, we completed two 3-month long observing campaigns of the CO J=2-1 rotational line using the Arizona Radio Observatory 10m Sub-millimeter Telescope during 2016 and 2018-2019, and compared the results to visible magnitudes obtained at the same time. As the Centaur approached its 2019 perihelion, the quiescent dust coma grew $\sim$45\% in brightness, while it is unclear whether the quiescent CO production rate also increased. A doubling of the CO production rate on 2016 Feb 28.6 UT did not trigger an outburst nor a rise in dust production for at least 10 days. Similarly, two dust outbursts occurred in 2018 while CO production continued at quiescent rates. Two other dust outbursts may show gas involvement. The data indicate that CO- and dust-outbursts are not always well-correlated. This may be explained if CO is not always substantially incorporated with the dust component in the nucleus, or if CO is primarily released through a porous material. Additionally, other minor volatiles or physical processes may help generate dust outbursts.

\end{abstract}


\newpage
\section{Introduction} 

Comets are icy bodies that contain well-preserved material from the early stages of the solar system formation. They formed at large heliocentric distances where temperatures were low enough for many ices to condense. Their nuclei contain trapped volatiles, refractory grains and rock. When they get close enough to the Sun, ices sublimate and release gas and dust grains that form envelopes known as comae. Some of these grains have frozen ices on them, which may then sublimate once immersed in the solar radiation field. Measuring the composition and structure of cometary objects, and how they become active, provides important observational constraints to models of the early solar system \citep{mum11,mandt15,fulle19}. 

Comets were historically classified into two broad categories based on their orbital periods: long-period (P$_{orb} >$ 200 years) and short-period comets. Long-period comets are distributed with random inclinations and large distances out to the Oort Cloud, whereas short-period comets tend to be found closer in and near the ecliptic, due to originating from the Trans-Neptunian region of the Kuiper Belt. Since orbital interactions with the giant planets, especially Jupiter, play a large role in many comets' dynamical evolution, the Tisserand parameter  ($T_j=a_J/a+2\sqrt{a/a_j(1-e^2)}\cos~i$) is also used to categorize comets \citep{Lev94}.  Many cometary objects whose orbits are Jupiter-controlled undergo a transitional Centaur phase and eventually become ``Jupiter Family'' comets (JFCs) and have  2 $< T_J <$ 3 \citep{lev97, done15, sarid19}.

29P/Schwassmann-Wachmann (hereafter 29P/SW1) is a remarkable cometary object. It temporarily occupies a nearly circular orbit, slightly larger than Jupiter's, with an eccentricity $e$ = 0.043, semi-major axis $a=$ 6.02 au, and inclination $i=$9.36 degrees, according to the Minor Planet Center (epoch 2019 March 18.0). 

Originally classified as a short-period comet ($P_{orbital}$ = 14.6 years), dynamical studies show that 29P/SW1 is a Centaur (with $T_J$=2.985) that probably only recently transferred to its current location, a newly discovered ``Gateway" transitional orbit to the JFC population, and may eventually become the brightest comet seen by humankind \citep{sarid19}. Thermal modeling of Spitzer-IRAC photometry predicts that the nucleus of 29P/SW1 is much larger than a typical comet, with a radius of 32 $\pm$ 2 km \citep{Schambeau2018}, and is by far the largest of the four objects known to be occupying the Gateway orbit. 29P/SW1's nucleus is probably larger than all JFCs and in a 1999 survey, its absolute nuclear magnitude was found to be brighter than all other Jupiter Family comets measured to date \citep{Fernandez1999JFC}.

29P/SW1 is one of $\sim$ 30 known active Centaurs and is legendary for having a continuously present dust coma superimposed with explosive outbursts several times a year (See Table \ref{tab:visual} and references on footnote $a$). The vast majority of its activity is documented with visible magnitudes, and most of the light at this wavelength comes from the reflection and scattering of sunlight off the dust particles in the coma, with smaller contributions from some ions and radicals \citep{cochrancochran91}. While small-scale changes in the lightcurve may be caused by nucleus rotation, large-scale changes in 29P/SW1's visible magnitude are probably largely due to changes in quantity or types of dust production output, including explosive outbursts.  Longterm tracking of 29P/SW1 with visible lightcurves show there is a slight preference for such outbursts to occur between perihelion and aphelion \citep{hughes1975,cabot96, krisandova14}, with an average of about seven per year over the most recent orbit \citep{trigo08, trigo10}. It is intriguing to consider how much more dramatic its behavior may become a few thousand years from now if its orbit takes it significantly closer to the Sun to become a JFC \citep{sarid19}.

We consider a dust outburst to be a brightening event of at least 1 magnitude of the nuclear magnitude (typically the central 5-10 arcsec part of the coma) that takes place within a few hours to a day. Brightness variations occur on smaller scales, but this threshold definition for an outburst is consistent with the analysis of many other published datasets  \citep{richter1941,roemer1958,trigo08,trigo10,miles16out,Schambeau2019}. The dust coma can also present morphological changes as an outburst progresses  \citep{schambeau17} and outbursts are documented at longer wavelengths (e.g., \citet{hosek13}. Not a great deal is known about the outburst characteristics, and there is very little  observational evidence for possible triggers.

\begin{table}
\centering
\caption{Typical visual magnitudes of 29P/SW1} 
\label{tab:visual}
\begin{tabular}{||l l l ||} 
 \hline
 {\bf Quantity} & {\bf Quiescent}  & {\bf Outbursting}    \\
 \hline\hline
$m_{v}$\footnote{Quiescent and outbursting values of visible magnitude reported for 29P, such as  \citet{roemer1958,roe62,whipple80, jew90, cochran91,trigo08, ferrin10, miles16out, Schambeau2018}} and this paper. & $\sim$ 17  & 16 -- 10 \\
$m_{helio}$\footnote{$m_{helio}$ is the apparent magnitude, $m_{v}$,  corrected for geocentric distance with equation 1, and is also referred to as $m(1,r,\theta)$,}  
& $\sim$ 13  & 12 -- 6 \\

 \hline

\end{tabular}
\end{table}

The ever-present dust coma is also remarkable, because at $\sim$ 6 au from the Sun, its nucleus is not warm enough to support much water-ice sublimation, the main source of activity for most comets, which usually becomes dominant when within $\sim$ 3 au of the Sun. Thus, 29P/SW1's coma must be generated by a different mechanism, as with other distantly active comets \citep{womack17}. Many alternatives have been considered for distant activity, including sublimation of a cosmogonically abundant low-condensation temperature ice, an amorphous-to-crystalline phase change of water-ice, HCN polymerization, cryovolcanism or meteoroid impacts \citep{prialnikbarnun90,sen94,enz97,gronkowski04,miles16cryo,Schambeau2019}.

\begin{table}
\centering
\caption{Measured and derived properties of dust and gas in 29P/SW1 coma }  \label{tab:dustgas}
\begin{tabular}{|l r r r r|}
 \hline
 {\bf Species} & {\bf Production rate} & {\bf mass loss rate\footnote{Gaseous mass loss rates were calculated using production rates and appropriate atomic mass units.}
} & {\bf Expansion speed} & {\bf Reference} \\
   & 10$^{27}$ mol/s & (kg/s) & (km/s) & \\
 \hline
Dust  & - & 430 -- 4700  & 0.05 -- 0.15 & [1,2,3,4,5,6,14] \\
CO & 10 - 70 & 460 -- 3200  & 0.20 -- 0.50 & [7,13]\\
H$_2$O & 6.3  & 188  & -- & [8] \\
N$_2$ (N$_2^+$)\footnote{N$_2$ production rate is inferred from CO$^+$, N$_2^+$ and CO measurements \citep{womack17}.} & 0.39 & 17 & -- & [7,9] \\
HCN (CN)\footnote{HCN production rate is inferred from Q(CN).} & 0.008   & 0.3  & --& [10] \\
H$_2$CO & 0.1 &$<$5 & --&  [12]\\
CS & $<$0.21 & $<$15 & --&  [12]\\
CO$_2$ & $<$0.35   & $<$25   & --& [8]\\
C$_2$H$_6$ & $<$0.57   & $<$28   & --&  [11]\\
CH$_3$OH & $<$0.55 & $<$29 & --& [12]\\
CH$_4$ & $<$1.3 & $<$34 & --& [11]\\
H$_2$S & $<$1 & $<$56 & --&  [12]\\
HC$_3$N & $<$0.95 & $<$81 & --&  [12]\\
C$_2$H$_2$ & $<$2.7 & $<$117 & --& [11]\\
NH$_3$ & $<$10.9 & $<$309 & --& [11]\\

\hline

\end{tabular}
\flushleft
[1] \citet{ful92}, [2] \cite{ivanova11}, [3] \cite{shi14}, [4] \cite{gun02}, [5] \cite{fulle98}, [6] \cite{fel96}, [7] \cite{womack17}, [8] \cite{oot12},  [9] \cite{ivanova16},  [10] \cite{cochrancochran91}, [11] \cite{pag13}, [12] \cite{biver97phd}, [13] \cite{biv99}, [14] \citep{Schambeau2018}
\\
\endflushleft

\end{table}

 A strong candidate for involvement with the activity is CO outgassing, since its emission was detected in 29P/SW1 over 25 years ago and has been measured on numerous occasions at millimeter- and infrared-wavelengths with high enough production rates to support lift off of the observed dust coma \citep{sen94,cro95,gun08,pag13}. In fact, CO has the highest production rate by far of all molecules measured in 29P/SW1. In Table \ref{tab:dustgas}, we list the measured production rates (or lowest significant limit obtained) for volatile species from the literature and then converted these to equivalent mass loss rates, which we calculated using  appropriate atomic mass units. The range of dust mass loss rates found in the literature is also given. Besides CO, other volatiles present in the coma are H$_2$O \citep{oot12}, CN  \citep{cochrancochran91}, CO$^+$ \citep{cochran80}, and N$_2^+$ \citep{kor08,ivanova16}. Water vapor was detected with production rates $\sim$ 20\% of typical CO \citep{oot12} and there is a preliminary report that high-resolution spectra of H$_2$O indicates that the emission is consistent with sublimation from icy grains in the coma \citep{Bockelee2014}. CN is a short-lived radical daughter species, probably created from photo-destruction of HCN in the coma. HCN emission was claimed to be detected in 29P/SW1 with relative abundances similar to that of Hale-Bopp at 6 au, and also with a spectral line profile consistent with sublimating from icy grains in the coma; however, no quantitative values have been published yet \citep{Bockelee2014}. Thus, the HCN values listed in Table \ref{tab:dustgas} were derived from CN measurement | see \citet{womack17} for more details. The amount of CO$^+$ detected during outbursting \citep{cochran80} and quiescent stages \citep{larson80} is consistent with forming from CO and not CO$_2$ \citep{ivanova19}. N$_2^+$ is detected in very small amounts, which was interpreted as evidence for N$_2$ being present only in small amounts relative to CO in the coma \citep{womack17}. Significant upper limits were obtained for the highly volatile CO$_2$ \citep{oot12}, and C$_2$H$_6$, CH$_4$, C$_2$H$_2$, NH$_3$, and CH$_3$OH \citep{pag13}, which indicate that all of these make much less of a contribution to the gas coma than CO (even when combined, see Table \ref{tab:dustgas}).

 With a perihelion distance of only $q=$ 5.76 au, the continuous presence of both dust and CO comae makes 29P/SW1 unique among the comet and Centaur populations. This may be typical of the beginning state of cometary activity for Centaurs as they progress toward a Jupiter Family comet orbit \citep{sarid19}. 
 
 There are many studies documenting 29P/SW1 quiescent and outbursting dust activity \citep{trigo08,hosek13,miles16out,Schambeau2019},  but very few include both dust and gas components measured at or about the same time. Analysis of optical spectra of 29P/SW1 obtained in 1990 indicated that the CO$^+$ and dust production were not strongly coupled during an outburst. The column density of the (3-0) CO$^{+}$ band at $\sim$ 4000\AA\ was observed to increase by a factor of 2.5, while the  continuum magnitude (dust), which was measured simultaneously at $\sim$ 4450\AA\ was unchanged. One interpretation was that this was evidence that the gas and dust was not entrained \citep{cochran91}, but this depended on an interpretation that CO$^+$ was a good tracer of the CO production rate. In comets, CO$^+$ can be produced by the photoionization of CO, but this process is slow and scales as $r^{-2}$, and is therefore ineffective at this large a distance from the Sun and is unlikely to be the source of most of the observed CO$^+$ \citep{cochrancochran91}. Instead, CO$^+$ column density and variability can be better explained by solar wind proton impact onto cometary CO, which is strongly dependent on solar wind particle velocities  \citep{cochran91,joc92,ivanova19}. Thus it appears that for 29P/SW1, CO$^+$ variation is not a straightforward proxy for CO behavior, and is a better tracer of solar wind particles at 29P/SW1's location for a given date. Thus, the reported variations of CO$^+$/dust continuum during outburst are likely not indicative of real changes of the gas/dust ratio without taking into account contributions from the solar wind.
 
Another study claimed a lack of correlation between the gas and dust production in 29P/SW1 evident in $Q(CO)$ and visible magnitudes \citep{biver01c}.  This conclusion is in contrast to a strong correlation that the author found for long-period comets C/1995 O1 (Hale-Bopp) and C/1997 J2 (Meunier-Dupouy).  However, it appears that the study for 29P/SW1 included all $Q(CO)$ and magnitude values, regardless of whether they were obtained during outbursting or quiescent stage. If the Centaur were having either a CO- or dust-outburst during any the \citet{biver01c} measurements, then this would significantly obscure any possible correlation. Comets Hale-Bopp and Meunier-Dupouy had far fewer outbursts, and so their data were highly likely obtained during when the comets were in a quiescent stage. To test for a correlation between the quiescent gas and dust production rate in this famously outbursting Centaur, one must compare known quiescent gas with quiescent dust production rates. Thus, although useful for showing that 29P/SW1 is a dust-poor comet compared to Hale-Bopp and Meunier-Dupouy, the results from \citep{biver01c} may not be sufficient to test the hypothesis of whether CO outgassing is correlated to the dust activity for either the quiescent or the outbursting stages. To address this, the $Q(CO)$ and magnitude values should be simultaneous and they should be taken when 29P/SW1 is in a quiescent or outbursting stage for both the gas and dust.
 
 There are also remarkably few detailed models attempting to describe the observational record \citep{enz97, kossacki13, wesolowski2018}.  Important observational constraints for testing models of 29P/SW1's behavior are missing: quantifying whether CO production is connected to dust outbursts, and if so, by how much. In order to assess this, we compiled secular lightcurves derived from visual brightness measurements obtained nearly at the same time as the CO millimeter-wavelength spectra and searched for correlations. In particular, it is widely assumed that CO outgassing is largely correlated with the dust outbursts in 29P/SW1, and we decided to test this hypothesis.

\label{introduction}

\section{Observations}

\subsection{Millimeter-wavelength spectra}

We monitored emission from the CO J=2-1 transition at 230.53799 GHz toward 29P/SW1 with the Arizona Radio Observatory 10-m Submillimeter Telescope (SMT) during 2016 February -- May and 2018 November -- 2019 January (see Table \ref{tab:CO}). The times listed in the table are at the end of the observing period for each day.

In both observing campaigns, the data were taken with the SMT 1.3 mm receiver with ALMA Band 6 sideband-separating mixers in dual polarization. The backend used for the observations consisted of a 500 channel (2 x 250) filterbank with a resolution of 250 kHz/channel in parallel mode. This frequency resolution corresponds to 0.325 km s$^{-1}$ per channel at the CO (2-1) frequency. The SMT beam diameter at this frequency was $\theta_B$ = 32$\arcsec$.
All of the scans were taken in beam-switching position mode with a throw of +2$\arcmin$ in azimuth, and the system temperatures remained between 200-400K. All the scans were obtained by integrating three minutes on the source and three minutes on the sky for subtraction. 

The position of the comet was checked periodically against the ephemeris position provided by JPL Horizons. Tracking was found to be within  $<$ 1$\arcsec$ RMS during both of the observing epochs. Additionally, pointing and focus was updated on planets and bright radio-source every six to ten scans.
\label{observations_spectra}

\begin{table}
\begin{center}
\caption{CO J=2-1 observations of 29P/SW1 using the ARO SMT 10-m telescope} 

 \label{tab:CO}
\begin{tabular}{|c c c c c c c|}
 \hline

 \bf{UT Date} & \bf{r} & \bf{$\Delta$} & \bf{T$_A^*$dv}  & \bf{$\Delta$v$_{FWHM}$} & \bf{$\delta$v} &\bf{Q} x10$^{28}$\\ [0.5ex] 
 & au & au & mK km s$^{-1}$ & km s$^{-1}$ & km s$^{-1}$&  mol s$^{-1}$ \\
 \hline
 2016 Feb 25.7 & 5.96 & 6.59 & 48$\pm$3  & 0.72 & -0.32 & *3.00$\pm$0.17 \\
 2016 Feb 26.7 & 5.96 & 6.57 & 59$\pm$5  & 0.89 & -0.39 & 3.67$\pm$0.28 \\
 2016 Feb 28.6 & 5.96 & 6.55 & 97$\pm$8  &  0.74 & -0.35 & 5.93$\pm$0.47 \\
 2016 Feb 29.7 & 5.96 & 6.54 & 55$\pm$5  & 0.74 & -0.48& 3.38$\pm$0.28 \\
 2016 Mar 01.8 & 5.96 & 6.52 & 49$\pm$3  & 0.83 & -0.36 & *3.05$\pm$0.20 \\
 2016 Mar 03.8 & 5.96 & 6.50 & 48$\pm$3  & 0.92 & -0.36 & *3.01$\pm$0.20 \\
 2016 Mar 21.7 & 5.95 & 6.24 & 71$\pm$3  & 0.75 & -0.26 & 4.36$\pm$0.22 \\
 2016 Mar 28.6 & 5.95 & 6.14 & 60$\pm$7  & 0.79 & -0.33 & 3.70$\pm$0.36 \\
 2016 Mar 31.6 & 5.95 & 6.09 & 59$\pm$5  & 0.68 & -0.38 & 3.62$\pm$0.28 \\
 2016 Apr 09.6 & 5.95 & 5.94 & 65$\pm$3  & 0.80 & -0.33 & 4.01$\pm$0.16 \\
 2016 Apr 15.6 & 5.95 & 5.85 & 42$\pm$3  & 0.70 & -0.29 & *2.62$\pm$0.18 \\
 2016 Apr 24.6 & 5.94 & 5.70 & 54$\pm$3  & 1.06 & -0.25& 3.34$\pm$0.15 \\
 2016 May 29.5 & 5.93 & 5.20 & 59$\pm$4  & 0.74 & -0.34 & 3.63$\pm$0.18 \\
 \hline
 2018 Nov 02.2 & 5.76 & 5.17 & 89$\pm$9  & 0.99 & -0.14 & *4.16$\pm$0.43 \\
 2018 Nov 10.2 & 5.76 & 5.28 & 86$\pm$6  & 1.13 & -0.30 & *4.06$\pm$0.29 \\
 2018 Nov 17.1 & 5.76 & 5.38 & 61$\pm$3  & 0.85 & -0.26 & *2.93$\pm$0.30 \\
 2018 Nov 24.2& 5.76 & 5.48 & 69$\pm$4  & 0.94 & -0.18 & *3.44$\pm$0.18 \\
 2018 Dec 12.2 & 5.76 & 5.75 & 78$\pm$4  & 0.87 & -0.30 & *4.06$\pm$0.20 \\
 2018 Dec 21.1 & 5.76 & 5.92 & 42$\pm$2  & 0.52 & -0.43 & *2.23$\pm$0.13 \\
 2019 Jan 08.1 & 5.76 & 6.19 & 79$\pm$3 & 0.85 & -0.34 & *4.40$\pm$0.17 \\
 2019 Jan 15.0 & 5.76 & 6.29 & 57$\pm$4  & 0.56 & -0.36 & *3.26$\pm$0.21 
\\
\hline

\end{tabular}
\end{center}
\footnotesize{*Q(CO) values marked with an asterisk are considered quiescent. (see \textit{Section 3}).}

\end{table}

\subsection{Visible magnitudes}
\label{sec:visible.magnitudes}

We used reports of 29P/SW1's apparent visual magnitude, $m_{v}$, to construct a secular lightcurve spanning the two time periods in which we monitored $Q(CO)$.  In order to minimize observer differences, we used magnitudes from experienced observers who were well-versed in how to obtain, reduce and report CCD magnitudes of comets  \citep{Larson1991} and Womack, M. et al. in prep. The data were obtained from the Lesia database of cometary  observations\footnote{http://lesia.obspm.fr/comets/index.php} and the Minor Planet Center Observation Database\footnote{ https://minorplanetcenter.net/db$\_$search} recorded with unfiltered CCDs (Table \ref{tab:visible1}). The magnitudes from the Lesia database of cometary observations were obtained with telescopes ranging between 0.2-m and 0.4-m in aperture diameter by the observers J.-F. Soulier, F. Kugel, J.-G.Bosch, J. Nicolas, T. Noel, and P. Ditz.   The observations from the MPC were obtained by TRAPPIST (0.6-m telescope in La Silla, Chile) and by J. Drummond  (0.35-m telescope at Possum Observatory in Gisborne, NZ).


\begin{longtable}{|c c c c c|}
\caption{Nuclear magnitudes,  $m_v$, of 29P/SW1 obtained from the Minor Planet Center and the Lesia database of cometary observations for both observing epochs. $\bf{D}$ is the telescope aperture diameter.}
\\
\hline
\bf{UT Date} & \bf{$\Delta$ (au)} & \bf{m$_v$}  & \bf{m(1,r,0)} & \bf{D(m)}\\ [0.5ex] 

\hline
\label{tab:visible1} & & & & \\
 2016 Feb 20.7 & 6.65 & 16.8 & 12.4 & 0.35 \\
 2016 Feb 24.4 & 6.63 & 17.2 & 12.8 & 0.60 \\
 2016 Mar 03.3 & 6.59 & 17.3 & 12.9 & 0.40  \\
 2016 Mar 08.3 & 6.51 & 17.3 & 12.9 & 0.40  \\
 2016 Mar 09.3 & 6.44 & 17.2 & 12.8 & 0.40  \\
 2016 Mar 10.3 & 6.43 & 17.3 & 12.9 & 0.40  \\
 2016 Mar 11.3 & 6.40 & 17.3 & 12.9 & 0.40  \\
 2016 Mar 12.3 & 6.39 & 17.4 & 13.0 & 0.40  \\
 2016 Mar 14.3 & 6.36 & 14.8 & 10.4 & 0.40  \\
 2016 Mar 16.3 & 6.33 & 15.0 & 10.6 & 0.40  \\
 2016 Mar 17.3 & 6.31 & 15.1 & 10.7 & 0.40  \\
 2016 Mar 18.3 & 6.30 & 15.3 & 10.9 & 0.40  \\
 2016 Mar 19.3 & 6.28 & 15.1 & 10.7 & 0.40  \\
 2016 Mar 21.3 & 6.25 & 15.7 & 11.3 & 0.40  \\
 2016 Mar 30.3 & 6.12 & 16.6 & 12.3 & 0.35 \\
 2016 Apr 04.3 & 6.04 & 16.6 & 12.3 & 0.32 \\
 2016 Apr 10.3 & 5.94 & 17.0 & 12.7 & 0.40 \\
 2016 Apr 12.3 & 5.91 & 17.1 & 12.8 & 0.40 \\
 2016 Apr 13.3 & 5.89 & 17.1 & 12.9 & 0.40 \\
 2016 Apr 14.3 & 5.88 & 17.1 & 12.9 & 0.40 \\
 2016 Apr 16.3 & 5.84 & 17.3 & 13.1 & 0.40 \\
 2016 Apr 19.1 & 5.79 & 17.4 & 13.2 & 0.30 \\
 2016 Apr 22.3 & 5.75 & 17.2 & 13.0 & 0.40 \\
 2016 Apr 23.3 & 5.73 & 16.9 & 12.7 & 0.40 \\
 2016 Apr 24.1 & 5.71 & 16.8 & 12.6 & 0.30 \\
 2016 Apr 25.3 & 5.70 & 17.3 & 13.1 & 0.40 \\
 2016 Apr 26.3 & 5.68 & 17.0 & 12.8 & 0.40 \\
 2016 Apr 27.1 & 5.67 & 17.4 & 13.2 & 0.40 \\
 2016 May 14.0 & 5.41 & 17.7 & 13.7 & 0.40 \\
 2016 May 17.0 & 5.36 & 17.4 & 13.4 & 0.20 \\
 2016 May 18.0 & 5.35 & 17.0 & 13.0 & 0.20 \\
 2016 May 20.0 & 5.32 & 16.7 & 12.7 & 0.20 \\
 2016 May 21.0 & 5.31 & 17.0 & 13.0 & 0.20 \\
 2016 May 22.0 & 5.30 & 16.8 & 12.9 & 0.20 \\
 2016 May 23.0 & 5.28 & 16.7 & 12.8 & 0.20 \\
 2016 May 24.0 & 5.27 & 16.9 & 13.0 & 0.20 \\
 2016 May 26.0 & 5.24 & 16.8 & 12.9 & 0.20 \\
 2016 May 28.0 & 5.22 & 16.1 & 12.2 & 0.20 \\
 2016 May 30.0 & 5.20 & 16.6 & 12.7 & 0.20 \\
 2016 May 31.0 & 5.18 & 16.6 & 12.7 & 0.20 \\
 2016 Jun 02.0 & 5.16 & 16.5 & 12.6 & 0.20 \\
 2016 Jun 03.0 & 5.15 & 16.7 & 12.9 & 0.20 \\
 2016 Jun 04.3 & 5.14 & 16.6 & 12.8 & 0.35 \\
 2016 Jun 05.0 & 5.13 & 16.5 & 12.7 & 0.20 \\
 2016 Jun 05.3 & 5.13 & 16.5 & 12.7 & 0.35 \\
 2016 Jun 06.0 & 5.12 & 16.3 & 12.5 & 0.20 \\
\hline
 2018 Nov 01.9 & 5.15 & 16.2 & 12.3 & 0.40 \\
 2018 Nov 02.7 & 5.17 & 16.1 & 12.2 & 0.30 \\
 2018 Nov 03.9 & 5.18 & 16.5 & 12.6 & 0.20 \\
 2018 Nov 04.9 & 5.19 & 16.4 & 12.5 & 0.40 \\
 2018 Nov 06.1 & 5.22 & 16.4 & 12.4 & 0.40 \\
 2018 Nov 08.9 & 5.25 & 16.6 & 12.6 & 0.40 \\
 2018 Nov 09.9 & 5.26 & 16.6 & 12.6 & 0.40 \\
 2018 Nov 10.9 & 5.28 & 16.3 & 12.3 & 0.40 \\
 2018 Nov 11.9 & 5.29 & 16.4 & 12.4 & 0.40 \\
 2018 Nov 13.0 & 5.32 & 16.4 & 12.4 & 0.40 \\
 2018 Nov 13.7 & 5.32 & 16.4 & 12.5 & 0.20 \\
 2018 Nov 13.9 & 5.32 & 16.6 & 12.6 & 0.40 \\
 2018 Nov 14.9 & 5.34 & 16.5 & 12.5 & 0.40 \\
 2018 Nov 15.7 & 5.35 & 16.1 & 12.1 & 0.20 \\
 2018 Nov 15.9 & 5.35 & 16.5 & 12.5 & 0.40 \\
 2018 Nov 16.8 & 5.37 & 16.2 & 12.2 & 0.30 \\
 2018 Nov 17.7 & 5.38 & 16.5 & 12.5 & 0.30 \\
 2018 Nov 18.7 & 5.40 & 16.5 & 12.4 & 0.30 \\
 2018 Nov 19.9 & 5.41 & 16.6 & 12.5 & 0.40 \\
 2018 Nov 20.9 & 5.43 & 16.7 & 12.6 & 0.40 \\
 2018 Nov 21.9 & 5.45 & 13.4 & 9.3 & 0.40 \\
 2018 Nov 22.7 & 5.46 & 13.6 & 9.5 & 0.20 \\
 2018 Nov 22.9 & 5.46 & 13.4 & 9.3 & 0.40 \\
 2018 Nov 23.9 & 5.48 & 13.5 & 9.4 & 0.40 \\
 2018 Nov 24.7 & 5.49 & 13.9 & 9.8 & 0.20 \\
 2018 Nov 24.9 & 5.49 & 13.9 & 9.8 & 0.40 \\
 2018 Nov 25.9 & 5.51 & 14.3 & 10.2 & 0.40 \\
 2018 Nov 27.0 & 5.54 & 14.7 & 10.6 & 0.40 \\
 2018 Nov 27.7 & 5.54 & 14.6 & 10.5 & 0.20 \\
 2018 Nov 28.0 & 5.56 & 15.0 & 10.9 & 0.40 \\
 2018 Nov 28.7 & 5.56 & 14.9 & 10.8 & 0.20 \\
 2018 Nov 28.8 & 5.56 & 15.9 & 11.7 & 0.41 \\
 2018 Nov 29.7 & 5.57 & 15.6 & 11.5 & 0.20 \\
 2018 Nov 30.0 & 5.59 & 15.7 & 11.6 & 0.40 \\
 2018 Dec 03.7 & 5.64 & 16.0 & 11.8 & 0.20 \\
 2018 Dec 04.0 & 5.65 & 16.2 & 12.0 & 0.40 \\
 2018 Dec 07.0 & 5.71 & 16.5 & 12.3 & 0.40 \\
 2018 Dec 08.0 & 5.72 & 16.3 & 12.2 & 0.40 \\
 2018 Dec 09.0 & 5.73 & 14.6 & 10.4 & 0.40 \\
 2018 Dec 10.0 & 5.75 & 14.8 & 10.6 & 0.40 \\
 2018 Dec 10.7 & 5.75 & 14.9 & 10.6 & 0.20 \\
 2018 Dec 11.7 & 5.76 & 15.2 & 11.0 & 0.20 \\
 2018 Dec 11.8 & 5.76 & 16.1 & 11.9 & 0.41 \\
 2018 Dec 13.0 & 5.80 & 15.7 & 11.5 & 0.40 \\
 2018 Dec 13.7 & 5.80 & 16.1 & 11.9 & 0.30 \\
 2018 Dec 14.8 & 5.81 & 16.2 & 12.0 & 0.20 \\
 2018 Dec 16.0 & 5.85 & 16.1 & 11.9 & 0.40 \\
 2018 Dec 17.7 & 5.87 & 16.4 & 12.2 & 0.20 \\
 2018 Dec 19.7 & 5.89 & 16.6 & 12.4 & 0.20 \\
 2018 Dec 20.7 & 5.91 & 16.8 & 12.6 & 0.20 \\
 2018 Dec 22.0 & 5.94 & 16.6 & 12.3 & 0.40 \\
 2018 Dec 23.0 & 5.95 & 17.1 & 12.8 & 0.40 \\
 2018 Dec 23.7 & 5.96 & 16.9 & 12.6 & 0.20 \\
 2018 Dec 24.0 & 5.97 & 16.7 & 12.4 & 0.40 \\
 2018 Dec 24.7 & 5.97 & 16.9 & 12.6 & 0.20 \\
 2018 Dec 25.7 & 5.99 & 16.9 & 12.6 & 0.20 \\
 2018 Dec 25.7 & 5.99 & 17.4 & 13.1 & 0.41 \\
 2018 Dec 26.0 & 6.01 & 17.0 & 12.7 & 0.40 \\
 2018 Dec 26.7 & 6.01 & 16.9 & 12.6 & 0.20 \\
 2018 Dec 26.7 & 6.01 & 17.3 & 13.0 & 0.20 \\
 2018 Dec 27.7 & 6.02 & 16.7 & 12.4 & 0.20 \\
 2018 Dec 28.7 & 6.04 & 16.5 & 12.2 & 0.20 \\
 2018 Dec 30.7 & 6.07 & 16.5 & 12.2 & 0.20 \\
 2018 Dec 31.7 & 6.08 & 16.6 & 12.3 & 0.20 \\
 2019 Jan 01.7 & 6.09 & 16.4 & 12.1 & 0.20 \\
 2019 Jan 02.7 & 6.11 & 16.3 & 12.0 & 0.20 \\
 2019 Jan 02.7 & 6.11 & 16.7 & 12.4 & 0.20 \\
 2019 Jan 03.7 & 6.12 & 16.0 & 11.7 & 0.20 \\
 2019 Jan 05.0 & 6.15 & 15.8 & 11.5 & 0.40 \\
 2019 Jan 06.0 & 6.16 & 15.8 & 11.5 & 0.40 \\
 2019 Jan 06.7 & 6.16 & 16.0 & 11.7 & 0.20 \\
 2019 Jan 07.7 & 6.18 & 15.9 & 11.6 & 0.20 \\
 2019 Jan 08.7 & 6.19 & 16.1 & 11.8 & 0.20 \\
 2019 Jan 09.0 & 6.21 & 16.3 & 12.0 & 0.40 \\
 2019 Jan 10.0 & 6.22 & 16.3 & 12.0 & 0.40 \\
 2019 Jan 10.7 & 6.22 & 16.3 & 12.0 & 0.20 \\
 2019 Jan 11.7 & 6.23 & 16.5 & 12.2 & 0.20 \\
 2019 Jan 12.7 & 6.25 & 16.7 & 12.4 & 0.20 \\
 2019 Jan 13.7 & 6.26 & 16.4 & 12.1 & 0.20 \\
 2019 Jan 14.7 & 6.27 & 16.8 & 12.5 & 0.20 \\
 2019 Jan 15.7 & 6.29 & 16.7 & 12.4 & 0.20 \\
 2019 Jan 21.7 & 6.36 & 16.8 & 12.5 & 0.20 \\
 2019 Jan 23.7 & 6.39 & 16.7 & 12.4 & 0.20 \\
 2019 Jan 24.7 & 6.40 & 16.7 & 12.4 & 0.20 \\
 2019 Jan 28.7 & 6.44 & 16.6 & 12.3 & 0.20 \\
 \hline

\end{longtable}

 We chose nuclear magnitudes measured with photometric apertures of 5 --7\arcsec~  centered on the nucleus, in contrast to much brighter ``total'' magnitudes, which generally encompass most of the visible coma.  Coma contamination typically occurs close to the nucleus; however, nuclear magnitudes are more likely than total magnitudes to pick up short-term changes in cometary activity.

We corrected the apparent visual magnitudes, $m_v$, for the Centaur's geocentric distance according to

\begin{equation}
m(1,r,\theta) = m_{v}-5log_{10}(\Delta)
\end{equation}
where $\Delta$ is the comet's geocentric distance and ranges from 5.2 - 6.6 au for the observing periods, and $m(1,r,\theta)$ is often referred to as the ``heliocentric magnitude,'' or $m_{helio}$. We also applied a phase correction to account for scattered light  using the phase function $\phi[\theta]$  normalized to 0$^{\circ}$ with 

\begin{equation}
m(1,r,0) = m(1,r,\theta) + 2.5log(\phi[\theta]),
\end{equation}

\noindent following the method of \citet{schleicher11}.\footnote{https://asteroid.lowell.edu/comet/dustphaseHM$\_$table.html}

\label{observations_magnitudes}

\section{Results} 

The CO millimeter-wavelength spectra and visible magnitudes presented here are similar to data reported elsewhere for the Centaur during its quiescent and outbursting stages  \citep{sen94, cro95, cabot96, biver97phd, fes01, gun03, gun08, trigo08, miles16out}.

\begin{figure}
\centering
\includegraphics[scale=0.4]{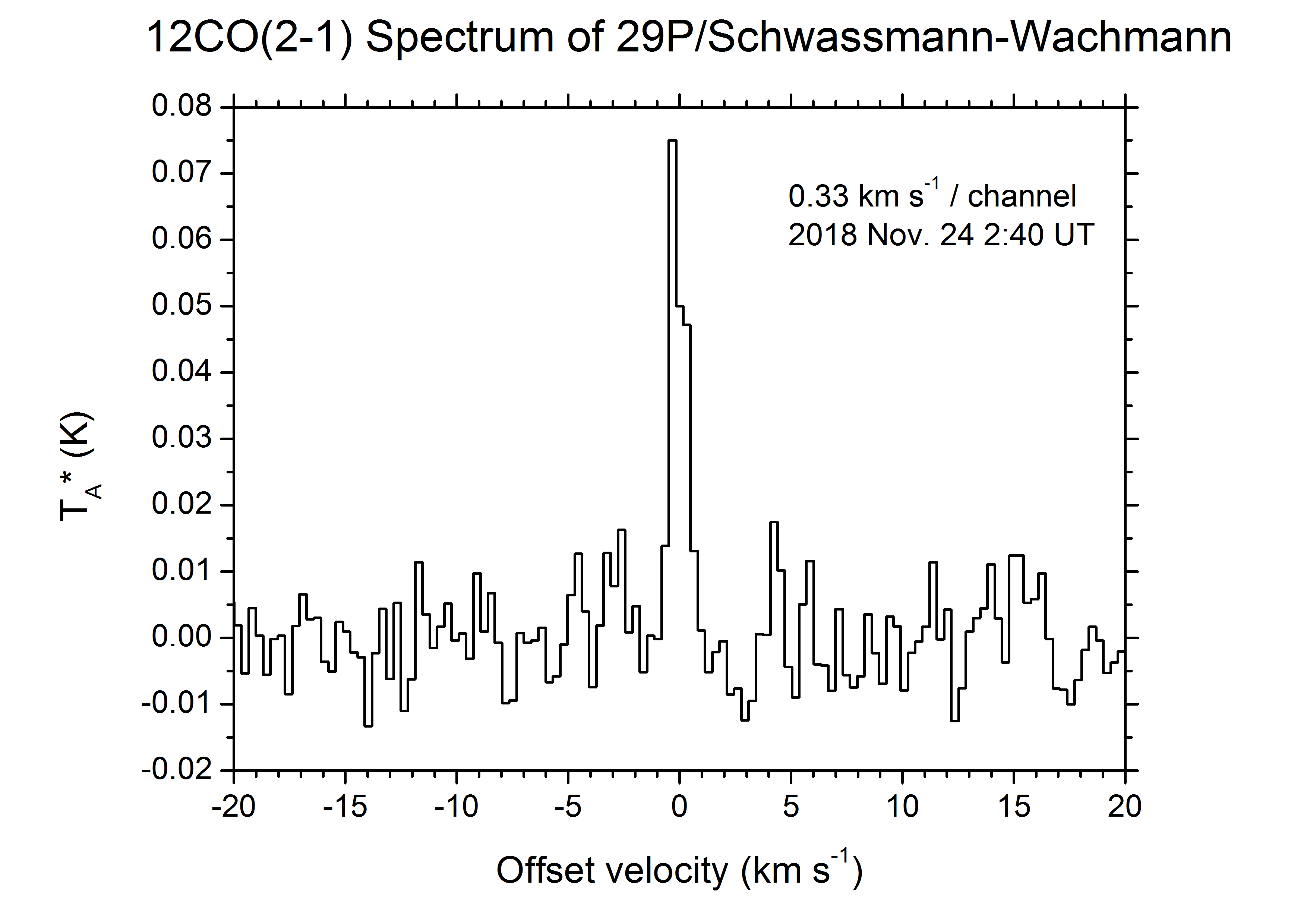}
\flushleft
\caption{A typical CO spectrum of comet 29P/SW1 obtained over $\sim$ 3 hours with the ARO 10-m SMT using 250 kHz (0.33 km s$^{-1}$) channel resolution.}
\label{fig:spectrum}
\end{figure}

Emission from the CO (2-1) rotational transition was readily detected in individual six-minute scans, and then were co-added with 3-6 scans to produce one spectrum for each day  (see Figure  \ref{fig:spectrum}). Using a Gaussian fit to the co-added spectrum the average full-width half-maximum linewidth for the CO emission was $\Delta$V$_{FWHM}$ = 0.87$\pm$ 0.33 km s$^{-1}$. Within the uncertainties, this is consistent with what is measured with other telescopes \citep{sen94,gun03}, considering that we did not fully resolve the line. The emission was blue-shifted by an average of $\delta v$ = -0.32 $\pm$ 0.16 km s$^{-1}$, in agreement with what is typically observed for CO emission in  cometary objects beyond 5 au \citep{womack17}. In order to derive CO column densities from the spectra, we used measured fluxes of the 2-1 line, and an excitation model that assumes fluorescence and collisional excitation following the modeling efforts of \citet{cro83} and \citet{biver97phd}. We assumed an optically thin gas, an excitation and rotational temperatures of 10K and a gas expansion velocity of 0.3 km s$^{-1}$. CO production rates were calculated from the column densities using a simple isotropic and constant radial outgassing model \citep{has57}. Line profile measurements and derived production rates are listed in Table \ref{tab:CO}.

\begin{figure}
    
\includegraphics[scale=0.47]{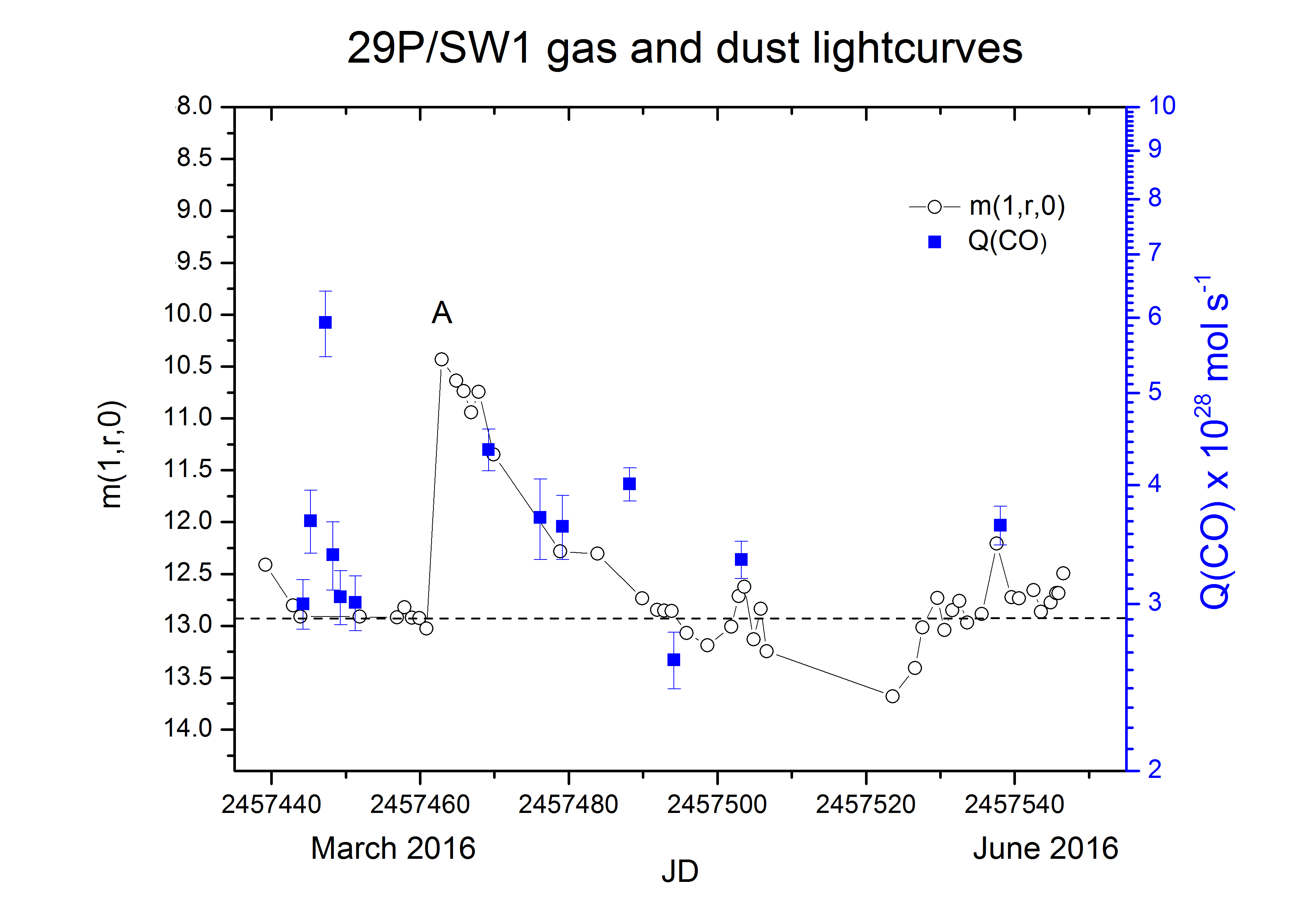}

\includegraphics[scale=0.47]{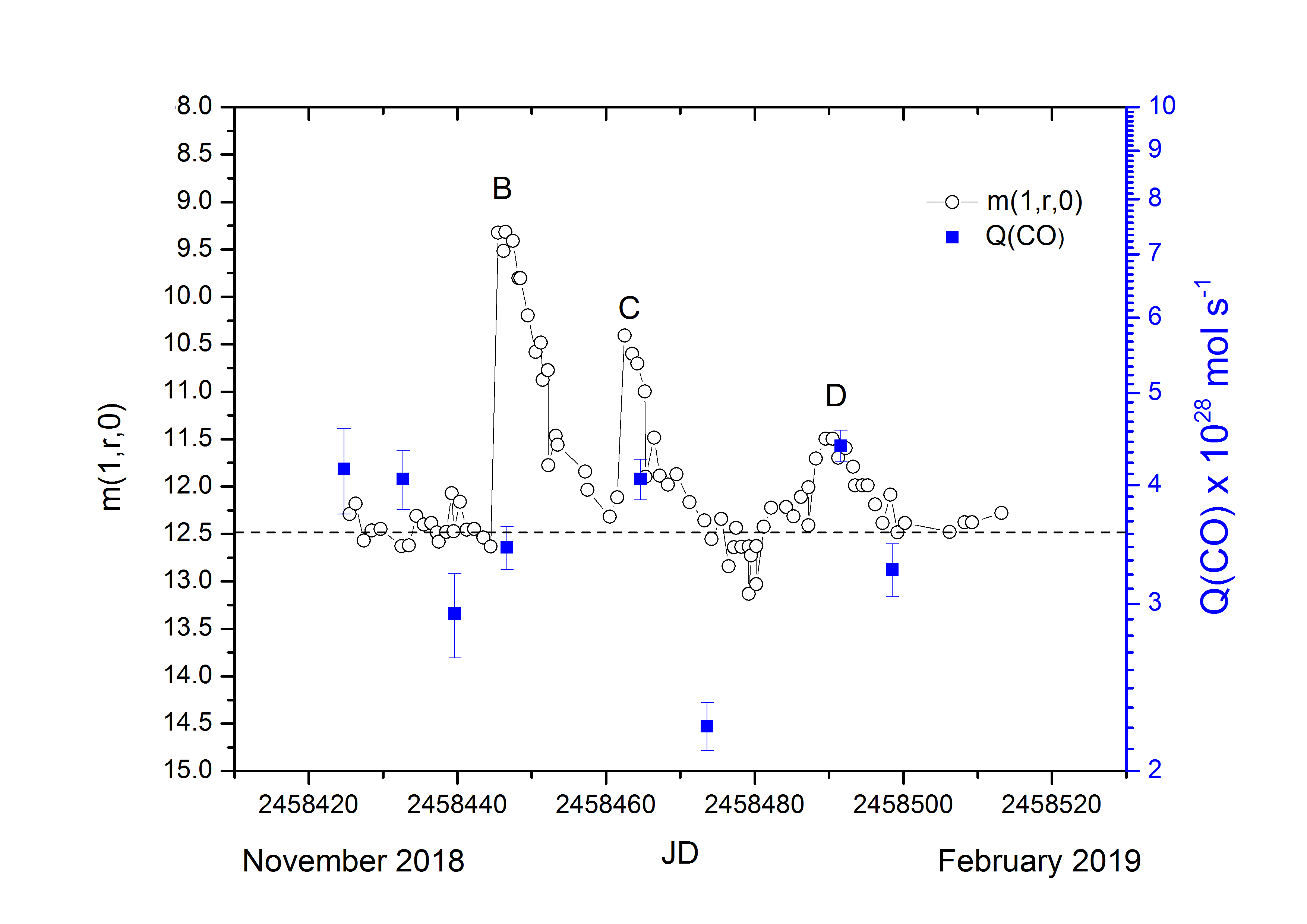}

\caption{29P/SW1 lightcurves of CO production rates (blue-filled squares) and visible magnitudes corrected for heliocentric distance and phase angle (open circles) during  2016 (top panel) and 2018-2019 (bottom panel). Each panel spans 120 days. One CO outburst was detected on 2016 February 28 which did not coincide with a dust outburst. Four dust outbursts were seen on 2016 March 14, 2018 November 22, 2018 December 09, and 2019 January 08 (marked as $A$, $B$, $C$, and $D$ respectively). The average quiescent stage values of $m(1,r,0)$ and $Q(CO)$ are indicated by a horizontal dashed line. The dust coma appeared $\sim$ 0.4 mag brighter in 2018-2019 when the Centaur was slightly ($\sim$ 0.2 au) closer to the Sun.}
   \label{fig:lightcurves}
   \end{figure}

\label{results}

Throughout our 21 days of CO observations, 29P/SW1 always produced at least $Q(CO)$ = 2-3 x 10$^{28}$ mol s$^{-1}$ with a few variations up to $\sim$ 4 x 10$^{28}$ mol s$^{-1}$ and a peak of $Q(CO)$ $\sim$ 6 x 10$^{28}$ mol s$^{-1}$  (see Table \ref{tab:CO} and Figure \ref{tab:dustgas}). The corrected visible magnitudes ranged from $m(1,r,0)$ = 9.3 to 13.7.

We calculated the average quiescent CO production rate and average quiescent heliocentric magnitude for each observing period using the data in Tables \ref{tab:CO} and \ref{tab:visible1}. For the 2016 epoch we define the quiescent CO stage as the apparent baseline level of CO production rate surrounding the CO outburst; that is values of $Q(CO)$ $<$ 3.2 x 10$^{28}$ mol s$^{-1}$. We did not include elevated CO values that appeared to coincide with increased dust production. With respect to 2018 - 2019, we did not definitely detect a compelling coincidence of elevated CO and dust production, so all measurements were used to determine the quiescent average value in this epoch. Quiescent values of $Q(CO)$ are marked with an asterisk in Table \ref{tab:CO}.

For 2016 the average quiescent CO production rate is $Q(CO)$ = (2.9 $\pm$ 0.2)x10$^{28}$ mol s$^{-1}$ and the average quiescent heliocentric magnitude is m(1,r,0) = 12.9 $\pm$ 0.2. For the 2018 - 2019 data the average quiecent values are $Q(CO)$ = (3.6 $\pm$ 0.7)x10$^{28}$ mol s$^{-1}$ and $m(1,r,0)$ = 12.5 $\pm$ 0.2. The uncertainties given in the averages are one standard deviation. 

In order to test the hypothesis that CO outbursts were correlated with dust outbursts we plotted the $Q(CO)$  and $m(1,r,0)$ values versus Julian date on the same graph (see Figure \ref{fig:lightcurves}). We aligned the average quiescent heliocentric magnitude and $Q(CO)$ values, which are designated with a horizontal dashed line in Figure \ref{fig:lightcurves}.

The observations recorded the evolution of one CO outburst over five days. Starting at 2016 February 25.7 the CO production rate doubled within 70 hours and then returned to the original quiescent production rate approximately three days later. Within the uncertainties, there is no measurable change to the CO line profile during the gas outburst. There are four short-lived dust outbursts of $\Delta$m $>$ 1 mag in the visible lightcurve, where $\Delta$m is the change in corrected visible magnitude $m(1,r,0)$ (Table \ref{tab:visible1}).  During the outbursts, the dust coma increased brightness within a few hours and decayed over a few days to weeks.

\section{Discussion} 

\label{discussion}

\subsection{Quiescent activity of gas and dust}
\label{quiescent}

Cometary activity is governed by the structure and composition of the nucleus, including how well the dust, ice and gas components are integrated \citep{pri04}. Activity is typically generated by the sublimation of water-ice, which is the largest icy constituent of most nuclei. However, at 29P/SW1's orbital distance water-ice sublimation off the nucleus surface is inefficient and instead, the volatile most likely responsible for driving the observed dust activity is considered to be CO based on its high production rates which match, and often exceed, the dust mass loss rate \citep{sen94}. 

As Table \ref{tab:dustgas} shows, no other volatile measured so far in 29P/SW1 competes with CO in terms of mass loss rates, including CO$_2$, CH$_4$, H$_2$CO, CH$_3$OH, N$_2$ or NH$_3$. O$_2$ is another possibly abundant molecule with high volatility ($T_{sub}  \sim$ 24K, \citet{yam85}), but it is very difficult to observe with telescopes \citep{LuspayKuti2018, Taquet2018} and no measurements of its abundance exist for 29P/SW1. Using mass spectroscopy O$_2$ was detected in the comae of 67P/Churyomov-Gerasimenko and Halley with amounts of $\sim$ 4\% relative to water \citep{bieler15, rubin15}. If O$_2$ is released with the same percentage to water in 29P/SW1's coma, then it would be present at Q(O$_2$) $\sim$ 2.4 x 10$^ {26}$ mol s$^{-1}$ (13 kg s$^{-1}$), placing it well below CO's production rate and on par with what we infer for N$_2$ (see Table \ref{tab:dustgas}). Of course, 67P and Halley were closer to the Sun when their measurements were made and they had more water-ice sublimation than 29P/SW1. Given the low sublimation temperature, if O$_2$ is present in 29P/SW1 then it may play a small role in the activity, but is not expected to rival CO. After considering the evidence for all likely competitors, we affirm that CO is the volatile most likely to be involved in generating the observed dust comae at optical and infrared wavelengths, but acknowledge that other molecules may be participating in less significant ways. 

Given the large role CO plays in 29P/SW1's gas coma, it is worth revisiting what observational clues we have to its behavior. At high spectral resolution, the CO line profile has a complicated structure with a strong blue-shifted velocity component, a weaker red-shifted component and a low-intensity skirted feature. A model containing a mix of sunward and nightside emission of CO, along with an extended icy grains coma matches this line profile well \citep{cro95,gun02,gun08}. The source of sunward emission is thought to be fairly close to the nucleus surface near the subsolar point to provide the continuous blue-shifted outgassing. This hypothesis is also supported by a recent updated model of dust outbursts for 29P \citep{Schambeau2018}. Although the spectra are relatively low resolution, the overall line shape of the data is blue-shifted by a small amount and is consistent with arising from a cold, slightly blue-shifted and asymmetric gas, as reported from other observers (see Figure \ref{fig:spectrum}). 

The narrow line profile of a cold gas may be a common feature in CO emission from many distantly active Centaurs and comets \citep{womack17}. In addition to Centaur 29P/SW1 ($\sim$ 6 au), this characteristic was documented in Centaurs 174P/Echeclus at $\sim$ 6 au \citep{wie17} and 95P/Chiron at 8.5 au \citep{wom99}, and comets C/1995 O1 (Hale-Bopp) beyond 4 au \citep{jew96, wom97}, C/2016 R2 (PanSTARRS) at $\sim$ 3 au \citep{wierzchos18, biver18}, and C/1997 J2 (Meunier-Dupouy) at 6.3 au \citep{biver01c}. Given the similarity of CO emission profiles in the Centaurs and distantly active comets observed so far, future observations, especially with simultaneous measurements of the dust coma, will be invaluable for constraining models of the volatile and refractory structure of the nuclei, and ultimately to their formation environment. 

Interestingly, the Centaur's quiescent dust coma measurably brightened over these three years: m(1,r,0) was 0.4 magnitude brighter during 2018-2019 (12.9 at $r$=5.76 au) than in 2016 (12.5 at $r$=5.95 au). This difference can be seen in Figure \ref{fig:lightcurves} and is consistent with the baseline level of dust activity increasing $\sim$ 45\% over the three years. During this time the CO quiescent level production rate increased by $\sim$24\%, which is within the measured uncertainties, and thus is also consistent with remaining steady. Given the uncertainties of the CO measurements, we cannot test the model that predicts that the dust and gas should change linearly during the quiescent stage  \citep{enz97}.  The Centaur was only about 3\%  closer to the Sun during 2018-2019, which is not large enough to attribute the increase solely to insolation change, which should vary as r$^{-2}$. Additional measurements are needed to confirm whether this change is part of a longterm brightening trend or just due to the variable nature of the dust production's quiescent stage (including possible contributions from nucleus rotation).

\subsection{CO and dust outbursts}
\label{COoutbursts}

Dust outbursts frequently brighten 29P/SW1's coma by several magnitudes within a few hours and generally resolve in under a week. Despite 29P/SW1's nearly circular path, the observing geometry (specifically the geocentric distance and phase angle) can contribute apparent changes of $\sim$1.5 magnitudes, but it takes several months, not days, to achieve such an effect, and we have already taken steps to correct this in Section \ref{sec:visible.magnitudes}. The telescope pointing errors were less than 2\arcsec ~ during the CO observations, which is not large enough to significantly affect the measured fluxes. Thus, the dust outbursts for which 29P/SW1 is so well known, and the rarely documented CO outbursts both require physical causes to explain their measured characteristics.

The nuclei of comets and Centaurs are assumed to be porous media containing a mixture of dust particles and volatile ices entrapped in adsorption sites \citep{klinger96}. Some models (eg. \citet{espin89,espin91,gun08}) assume the nucleus is composed substantially of amorphous water-ice containing trapped highly volatile gases, like CO or CO$_2$, whose release is triggered by phase change of crystallization of the amorphous water-ice (similar to models of Hale-Bopp activity at large distances, e.g. \citet{prialnik97}). At 29P/SW1’s distance from the Sun, the effective blackbody surface temperature is $\sim$ 120 K, which meets the energy threshold for the amorphous-crystalline change (See Equation 1 from \citet{womack17}). Once released, gases may travel different paths out of the nucleus depending on their volatilities, some perhaps starting with sublimation from below the surface. In addition, other surface processes may trigger erosion and mass loss, which may lead to new release of gas without simultaneous release of dust and vice versa \citep{enz97, meech04, pri04, steckloffjacobson2016}. In addition, a recent updated model of dust outbursts in 29P/SW1 is tied to the emission of CO from the sunward side \citep{Schambeau2018}. 

As Table \ref{tab:CO}  shows, there was no measurable change in either the Doppler shift or FWHM linewidth during the 2016 CO outburst. Thus, there is no observational evidence that the doubling of CO production arose from a change in the outgassing mechanism responsible for the quiescent stage (described in section \ref{quiescent}).  We point out that \citet{biver97phd} also reported a  factor of $\sim$2 increase in $Q(CO)$ in 29P/SW1 that decayed in 2 -- 3 days after the peak. Those data were obtained during 1995 November 15-19 with the IRAM 30-m telescope. Thus, there are two documented cases of development of a CO outburst in 29P/SW1, and both lasted $\sim$ 4-5 days. 

The 2-3 day return to quiescent level after the CO outburst is consistent with most of the molecules traveling out of the radio telescope beam after a few days. For example, consider that the SMT FWHM beamwidth at 230 GHz is 32\arcsec, which corresponds to a projected radius of 63,500 km at the comet's distance on 2018 November 24 ($\Delta$ = 5.4 au).  
If we assume a CO expansion velocity of 0.3 km s$^{-1}$ (consistent with all modeling of CO mm-wavelength spectra), then we estimate that molecules traveling parallel to the plane of the sky (zero blue-shift to the geocentric velocity) would ``leave the beam'' in about 2.4 days (t=x/v=6.35 x 10$^4$ km / 0.3 km s$^{-1}$ = 2.4 days), and more quickly with the smaller IRAM beam. Molecules traveling out of the plane of the sky would remain in the beam a little longer since their tangential velocity is reduced. Thus, the total $\sim$ 4-5 day duration of the CO outbursts seen with the ARO SMT in 2016 (this paper) and with IRAM 30-m in 1995 \citep{biver97phd} are consistent with a very short-term (few hours) release of CO molecules from the nucleus and then most of the molecules moving out of the telescope beam. CO's lifetime against photo-destruction is very long at this heliocentric distance and does not significantly reduce the CO coma size in a few days. Additional observations of CO outbursts in 29P/SW1 with higher temporal, spectral and spatial resolution would significantly constrain models describing processes that affect CO emission at large distances from the Sun.

CO outgassing is often assumed to play an important role in the dust outbursts, largely due to its high production rate and the adequacy of some models to explain the quiescent activity \citep{cowan82,enz97}. In order to test the hypothesis that CO production is connected to the dust outbursts, we constructed the $Q(CO)$ and visible magnitude lightcurves.  Intriguingly, the 2016 CO gas outburst was not immediately accompanied by an increase in dust production. The dust coma maintained a nearly constant quiescent magnitude of $m(1,r,0)$ = 12.9 $\pm$ 0.2  for at least the ten days after the CO outburst (Figure \ref{fig:lightcurves}). Thus, either the CO outburst led to little-to-no increase in dust production, or it created a dust outburst after a $\sim$ 10 day delay.  Furthermore, our analysis also indicates that CO outbursts are not required to generate  dust outbursts (see outburst $\bf B$ in Figure \ref{fig:lightcurves}). The previously mentioned 1995 CO outburst reported by \citet{biver97phd} did not have simultaneous measurements of the dust coma; thus, these 2016 data are the only known set of CO outburst in 29P/SW1 with quasi-simultaneous information about the dust.

The two CO gas outburst patterns are similar to some of the visible outburst patterns in that they have a rapid rise with a slow decline.  Both CO gas outbursts appeared to decay to quiescent levels a day or two more quickly (at least) than typical dust outbursts. When documented with lightcurves, dust outbursts appear to return to quiescent values in 3-5 days, sometimes longer \citep{trigo08,trigo10,miles16out}. The longer decay time for dust can be explained by the dust having a lower expansion velocity than the gas  \citep{Schambeau2018}. 

There are four noticeable dust outbursts during the observing campaign. Two occurred without a substantial increase in CO production (labeled $\bf{B}$ and $\bf{C}$ in Figure \ref{fig:lightcurves}), and two may have coincided with a rise in CO  ($\bf{A}$ and $\bf{D}$).    Here we describe and discuss all four in detail.
On 2016 March 14 29P/SW1 underwent a $\sim$ 3 magnitude dust outburst (labeled $\bf A$ in Figure \ref{fig:lightcurves}), which corresponds to a $\sim$ 16-fold increase in  brightness.  After peak brightness, the coma steadily dimmed over weeks to its quiescent value, a decay time much longer than the other dust outbursts in this paper and reported elsewhere (e.g., \citet{trigo08,miles16out}). We also consider that the lightcurve pattern for $\bf{A}$ may be due to two or more outbursts occurring within a few days of each other.

The second outburst, $\bf{B}$, brightened by $\sim$ 3.3 magnitudes (a 20-fold increase) on 2018 Nov 22, which it maintained for a few days before decaying. As Figure \ref{fig:lightcurves} shows, 52 hours after the start of this outburst, when the dust outburst was still at its highest point, the CO production rate was at the quiescent value. Therefore, dust outburst $\bf{B}$ appears to have been triggered, and even maintained over $\sim$ 2 days, without an increase in the CO production rate.

Outburst $\bf{C}$ began on 2018 December 09, increased brightness by $\Delta$m = 2.5 magnitude, and 2.5 days later was accompanied by a slightly elevated CO production rate.  Hypothetically, if the dust outburst mechanism for $\bf{B}$ and $\bf{C}$ were tied to a significant release of CO, then such a CO outburst must have resolved itself within 1-2 days, i.e., more quickly than the observed CO outburst. This is further indication of either no CO involvement with a dust outburst, or perhaps a scenario where two other short-lived CO outbursts occurred when we were not collecting data with the SMT. Regardless, whatever the mechanism was behind the $\bf{B}$ and $\bf{C}$ outbursts, they did not lead to a multiple day long increase of the CO emission. 

The smallest dust outburst, labeled {\bf D}, which peaked on 2019 January 08 with $\Delta$m=1.1 magnitude, does not have the same profile as the first three outbursts. Instead of a rapid brightening over a matter of hours, it appears to more gradually brighten over several days and then decay somewhat symmetrically, possibly indicating a different physical process at work. Also, outburst {\bf D}
coincides with a similar increase in $Q(CO)$, and there is some indication that both the gas and dust returned to quiescent levels a week later. Unfortunately, there are not enough measurements of the CO emission to further explore the gas relationship to the dust for this outburst. This is further demonstration of the need to obtain more time-series observations of CO spectra along with measurements of the dust coma.

The data presented in this paper show that only two of the dust outbursts, {\bf A} and {\bf D}, may have been associated with an increased CO production rate. The paucity of CO measurements around the start of the dust outburst $\bf{A}$ makes drawing conclusions difficult, but there is some evidence for CO involvement, as the $Q(CO)$ and visible magnitude declines at similar rates during the weeks after the outburst, particularly on March 21 --31. This similar apparent rate of decline could be also consistent with an extended source of CO from the coma grains at this time (e.g., \citet{gun03}). There was no such evidence for an extended source in the CO outburst that occurred a few weeks earlier. In contrast, the CO outburst of early 2016 did not lead to an increased dust production for at least 10 days, and neither dust outburst {\bf B} or {\bf C} were associated with an increased CO production rate. Thus, CO and dust outbursts are not always correlated, and we may be observing CO arising from both the nucleus and an extended source in the coma.

The lack of correlation between the CO- and dust-outbursts in the data may be explained if CO is not substantially incorporated in a material with the dust component in the nucleus, or if the surface is regularly interrupted. Other possibilities to explain the separate dramatic release of large quantities of gas and dust is if the CO is primarily released through a highly porous material, or if another relatively minor volatile plays a large role in the dust outbursts.

We briefly consider other volatile candidates that might be involved in triggering the dust outbursts. A common candidate proposed for activity in many distantly active comets is CO$_2$ \citep{oot12,bauer15,womack17}. In particular, \citet{gun08} suggest that CO$_2$ (with its higher atomic weight) might be  mass-loading with CO in 29P/SW1 and slowing down its exit into the coma. However, using simultaneous spectroscopy from the Akari infrared telescope, CO$_2$ has a very low measured upper limit of $Q(CO_2)$ $<$ 3.5 x10$^{26}$ mol s$^{-1}$ in 29P/SW1, corresponding to a mass loss rate of dM/dt $<$25 kg s$^{-1}$ and a  mixing ratio of $Q(CO_2)/Q(CO) <$ 0.01 \citep{oot12}.  Measurements of the aggregate emission from CO$_2$+CO at 4.5 $\mu$m were made with photometry that are also consistent with very little CO$_2$ in this Centaur (Harrington Pinto, O. et al. in prep.) This is well below the predicted value CO$_2$ in \citet{gun08} and thus the data are not consistent with significant mass-loading of CO$_2$ with CO, or with the dust outbursts, unless only a very minimal amount ($<$ 25 kg s$^{-1}$) is needed. 

As for other possible candidates for outbursting activity, C$_2$H$_2$ and NH$_3$ also have relatively low sublimation temperatures (57K and 78K respectively, \citet{yam85}) and are detected in other comets with low abundances. They have not been identified in 29P/SW1, but their upper limits are not very constraining: $Q(C_2H_2) <$ 117 kg s$^{-1}$ and $Q(NH_3) <$ 309 kg s$^{-1}$ (Table \ref{tab:dustgas}), so efforts to observe these two molecules may be fruitful. Other cosmogonically abundant species with low sublimation temperatures already have very low upper limits placed on them, and are unlikely to significantly compete with CO in the coma (see Table \ref{tab:dustgas}).

The dust outbursts in the lightcurve demonstrate an important point about the sporadic nature of the outbursts. 29P/SW1's rotation period is in great disagreement between different authors with values ranging between 14 hours and 50 days \citep{meech93,whipple80,jew90,trigo10,ivanova12,Schambeau2018}. This is a substantial problem, because an accurate estimate of 29P/SW1's rotation period is needed to model its activity in more detail. \citet{trigo10} partly favor a rotation period of $\sim$ 50 days because of the stated annual average of $\sim$ 7 dust outbursts/year, which corresponds to an average of one outburst every $\sim$ 50 days. However, this average time between outbursts carries a high dispersion: many $>$1 magnitude outbursts occur only 1-2 weeks apart (and not $\sim$ 50 days) as the data in Figure \ref{fig:lightcurves} show. Even the quiescent stage is highly variable over days to weeks which makes it difficult to establish the start and end times of dust outbursts. Furthermore, some dust outbursts may overlap in time, which may be the case for the outburst $A$. Thus, as \citet{trigo10} points out, the $\sim$ 50 day average obtained from an observed 7 outbursts/year may not be physically tied to the actual rotation rate of the nucleus.

In order to constrain the mechanisms driving the activity of this enigmatic Centaur, further observations of 29P/SW1 at all wavelengths with myriad techniques are strongly encouraged |  especially simultaneously. 29P/SW1 recently underwent perihelion in 2019 March, and it has now entered the epoch when its activity historically increases \citep{cabot96,krisandova14}. 29P/SW1's activity level increased significantly after its orbit changed due to a close interaction with Jupiter in 1975. In 2038, 29P/SW1 will have another important conjunction with Jupiter that is predicted to double its eccentricity \citep{sarid19}. 29P/SW1's activity level may very well be affected by this new orbit change and this is another motivation for closely monitoring the Centaur in the next few decades. Continuous longterm ($>$ 80 days) monitoring of the CO production rate, preferably with high spectral resolution, is needed to document any increases in $Q(CO)$ that lead to dust outbursts, and if so, the evolution and relationship of the two phenomena. Time-series monitoring of other molecular species will also be useful to constraining improved models of the dust outbursts. A world-wide observing campaign ({http://wirtanen.astro.umd.edu/29P/29P\_obs.shtml}) for 29P/SW1 was initiated in 2018 by M. Womack, G. Sarid and T. Farnham. Participation is strongly encouraged as it may help different teams coordinate observing runs and increase the chances of obtaining simultaneous measurements and maximizing scientific return.

\label{discussion_18_dust_outburst}

\section{Summary and Conclusions}

For years it has been hypothesized that CO outgassing stimulates the release of dust outbursts in 29P/SW1, as well as the quiescent dust coma. Comparison of the $Q(CO)$ and visible lightcurve data, an effective tracer of dust production in the Centaur, show that the relationship between outbursts of dust and gas is not always this straightforward. 

Throughout 21 days of CO observations over three years 29P/SW1's quiescent gas production rate changed from  $Q(CO)$ = (2.9$\pm$0.2)x10$^{28}$ mol s$^{-1}$ to (3.6$\pm$0.7)x10$^{28}$ mol s$^{-1}$. This is an increase of $\sim$ 24\%, but is also consistent with remaining steady within the uncertainties. The quiescent value of the heliocentric magnitude corrected for phase angle brightened by $\sim$ 45\% from  m(1,r,0) = 12.9 $\pm$ 0.2  to  12.5 $\pm$ 0.2, while the heliocentric distance decreased by only 4\%.

Several outbursts were recorded during the observing periods. On 2019 Feb 25.7 the CO production rate doubled within 70 hours and then returned to the original quiescent value approximately three days later. The CO emission line profile characteristics appeared typical of what is observed for CO in distant comae and showed no evidence for substantial deviation related to the outbursting mechanism. The CO increase was unmatched by a change in dust production for at least ten days, consistent with no significant dust involvement as a consequence of the CO outburst.

Four dust outbursts were recorded at visible wavelengths: two ({\bf B} on 2018 Nov 22 and {\bf C} on 2018 Dec 9) occurred without a measurable increase in CO production  and two ({\bf A} on 2016 Mar 14 and {\bf D} on 2019 Jan 8) coincided with increased CO amounts. The lack of a strong correlation of gas and dust behavior during all of the CO and dust outbursts shows that although the CO and dust production increases may be sometimes coincident, they are not always significantly entrained in 29P/SW1. The data are not consistent with the commonly held assumption about 29P/SW1 that increased CO outgassing must always be involved with the dust outbursts and may provide important observational constraints for cometary outbursting models.  

The dust outburst lightcurve shapes may hold additional clues to the mechanisms involved: outbursts {\bf A, B,} and {\bf C} had an asymmetric shape with a steep buildup and gradual decay, whereas outburst {\bf D} appeared more symmetric about the peak including a more gradual buildup and decay. Outburst {\bf D} (which also coincided with elevated CO production) appears to have been triggered with a less explosive process than the other three, and it is also the smallest outburst observed of this group. Additional coordinated observations of multiple species in the comae of 29P/SW1 are needed to further test models of the interplay of the gas and dust components in driving the quiescent activity and outbursts.

\section{Acknowledgements}

This material is based on work supported by the National Science Foundation under Grants No. AST-1615917 and AST-1945950 (MW as PI). KW was partially supported from the University of South Florida Duckwall Summer Research Fellowship.  The authors are grateful to an anonymous referee who provided a critical reading of the manuscript and helpful comments. The authors also thank the ARO 10m SMT staff, the Minor Planet Center, and the observers who contribute to the Lesia database of comet observations. The SMT is operated by the ARO, the Steward Observatory, and the University of Arizona, with support through the NSF University Radio Observatories program grant AST-1140030.

\bibliography{BigReferences_2019.bib}

\end{document}